\documentclass[useAMS,usenatbib]{mn2e}

\usepackage{amssymb}
\usepackage{aas_macros}
\usepackage{natbib}
\usepackage{times}
\usepackage[dvips]{graphicx}

\title[Formation of a large disk galaxy]{Forming a Large Disk Galaxy from a  z$<$1 Major Merger}
\author[F.Governato  et al. ]
{
\parbox[t]{\textwidth}{
 F.~Governato$^{1}$\thanks{E-mail:fabio@astro.washington.edu}, 
 C.~B.~Brook$^1$,
 A.~M.~Brooks$^{1,2}$,
 L.~Mayer$^{3}$,
 B.~Willman$^{4,5}$,
  P.~Jonsson$^6$,
  A.M.~Stilp$^1$,
  L.~Pope$^1$,
  C.~Christensen$^1$,
  J.~Wadsley$^7$,
  T.~Quinn$^1$}
   \vspace*{6pt} \\
$^1$ Department of Astronomy, University of Washington, Box 351580, Seattle, WA 98195, USA;\\
$^2$ California Institute of Technology, Theoretical Astrophysics, MC 130-33, Pasadena, CA 91125;\\
$^3$ University of Zurich \& ETH,  Zurich, Switzerland;\\
$^4$ Clay Fellow, Harvard-Smithsonian Center for Astrophysics, Cambridge, MA;\\
$^5$ Dept. of Physics and Astronomy, Haverford College, 371, Lancaster Ave Haverford PA, 19041 MA\\
$^6$ Institute of Particle Physics, University of California, Santa Cruz, CA 95064, USA\\
$^7$ Dept. of Physics and Astronomy, Mac Master University, Hamilton, ON, Canada
\vspace*{-0.2cm}}

\begin{document}

\date{Submitted to MNRAS}

\pagerange{\pageref{firstpage}--\pageref{lastpage}} \pubyear{2002}

\maketitle

\label{firstpage}

\begin{abstract}

  Using high resolution SPH simulations in a fully cosmological
  $\Lambda$CDM context we study the formation of a bright disk
  dominated galaxy that originates from a ``wet'' major merger at
  z$=$0.8. The progenitors of the disk galaxy are themselves disk
  galaxies that formed from early major mergers between galaxies with
  blue colors. A substantial thin  stellar disk grows rapidly following the
  last major merger and the present day properties of the final
  remnant are typical of early type spiral galaxies, with an $i$ band
  B/D $\sim$ 0.65, a disk scale length of 7.2 kpc, $g-r$ = 0.5 mag, an
  HI line width ($W_{20}$/2) of 238 km/sec and total magnitude $i$ =
  -22.4.  The key ingredients for the formation of a dominant stellar
  disk component after a major merger are:  i) substantial and rapid
  accretion of gas through cold flows followed at late times by
  cooling of gas from the hot phase, ii) supernova feedback that is
  able to partially suppress star formation during mergers and iii)
  relative fading of the spheroidal component. The gas
  fraction of the progenitors' disks does not exceed 25\% at z$<$3,
  emphasizing that the continuous supply of gas from the local
  environment plays a major role in the regrowth of disks and in
  keeping the galaxies blue.  The results of this simulation alleviate
  the problem posed for the existence of disk galaxies by the high
  likelihood of interactions and mergers for galaxy sized halos at
  relatively low z.

\end{abstract}


\section{Introduction}

Within the $\Lambda$CDM framework the build up of galaxies and their
parent dark matter (DM) halos occurs through a series of mergers and
accretion of more diffuse matter \citep{frenk85, millennium05}. The
classic models of galaxy formation
\citep{whiterees78,fall83,whitefrenk91} and subsequent work
\citep{dalcanton97,mo98,silk01,bower06,zheng07,somerville08} assume
that gas infalls inside the parent dark matter halo and subsequently
cools to temperatures low enough to fragment and form stars. The
morphology of galaxies and the fraction of stars in their disk and
spheroidal components are set by a combination of competing physical
processes, as numerical simulations have shown that violent relaxation
in mergers between similar size galaxies can destroy or dynamically
heat their stellar disks and turn them into spheroids
\citep{barneshernquist96}. On the other hand, remaining and
subsequently accreted gas can regrow a disk \citep{baugh96,
  steinmetznavarro02} if given enough time. It is then a prediction of
hierarchical models that the morphology of a galaxy is not assigned
``ab initio'' but it might change often over cosmic times, as the
spheroidal to disk light ratio changes back and forth.  Observed
merger remnants have disky isophotes \citep{rothberg04} or evidence
for young disk components \citep{mcdermid06}, hinting at disk regrowth
and supporting the above scenario.

Observationally, major mergers are observed to be common in the
redshift range 0 -- 1. Their number density has been estimated to be
of the order of 10$^{-(3-4)}$h$^{-1}$Mpc$^{-3}$ Gyr$^{-1}$,
\citep{lin08,jogee08}, possibly increasing with redshift. As numerical work
suggests that a consistent supply of gas is important to regrow a disk
component \citep{robertson06}, it is relevant that observational
evidence for mergers between actively star forming galaxies has also
emerged, showing that at redshift $\sim$1, 70\% of the merging pairs
are between blue  (or ``wet'', defined as rest frame g-r $<$0.65) pairs; however the fraction
of blue and  likely gas rich mergers decreases  by
the present time \citep{lin08}.
 
Numerical work has also highlighted a possible problem for the
existence of disk galaxies at low redshift, showing that interactions,
mergers and accretion events are common at every redshift for L$\star$
galaxy sized halos in $\Lambda$CDM models \citep{maller06,stewart08}
and potentially destructive for disks
\citep{toth92,stelios07,bullock08,purcell08}.  This makes it
potentially difficult to reconcile the observed present day population
of disk dominated galaxies with $\Lambda$CDM, if a quiet merging
history is essential for their existence. But do disk galaxies really
require a quiet merging history to be observed as such at the present
time? What are the time scales of disk growth and destruction?  How
quickly do disk galaxies regrow after a merger?  Theoretical models
also highlight the difficulties in predicting the outcomes of galaxy
mergers and the subsequent regrowth of  stellar  disks. These
difficulties arise from several factors. Given the decoupling of the
baryonic cores from the parent halos it is difficult to predict robust
merging rates of galaxies (as compared to their DM halos) unless
cosmological simulations with hydrodynamics are used \citep{maller06}.
Also, numerical simulations have shown how the bulge to disk ratio
(B/D) resulting from a major merger might depend on the orbital
parameters, the internal spin of the progenitors, the gas fraction of
the parent disks and the efficiency of SN feedback
\citep{barneshernquist96,cox06,scannapieco08}.

Recent observational and theoretical work has pointed out ways to
alleviate the possible problem of a high merger rate for the survival
of disk galaxies. \cite{hopkins08} showed that angular momentum loss
of the gas component is not necessarily catastrophic even in 1:1
mergers. Disks can survive or rapidly regrow, provided that the gas
fraction in the disks of the progenitors is high
\citep{robertson06,robertson08,bullock08}.  If feedback is able to
suppress star formation during the merger event
\citep{brook04,robertson06,zavala08,g07} the existing cold
gas can settle on a new disk plane and start regrowing a stellar disk.

Hydrodynamical simulations indeed show that cold flows (cold gas that
flows rapidly to the center of galaxies from filamentary structures
around halos \citep{keres05,dekel08}) play a major role in the build
up of disks in galaxies \citep{brooks08}. Gas accreted through cold
flows arrives to the central  stellar disk on a time scale a few
Gyrs shorter than gas that is first shocked to the virial temperature
of the host halo and then cools onto the disk, leading not only to
early disk star formation, but the creation of a large reservoir of
cold gas.  No matter its origin, late infalling gas would likely have
a higher angular momentum content than material accreted at earlier
times \citep{quinn92} and gas in the merging disks would acquire a
coherent spin set by the orbital parameters of the binary system and
the internal spins of the parent galaxies.  A feedback$+$cold flows
model is particularly attractive as the mechanism for the
survival/regrowth of  gas (and then stellar) disks as it provides
a natural explanation to the fact that more massive galaxies tend to
have large B/D ratios \citep{benson07,graham08}, while smaller
galaxies are likely disk dominated.  At large galaxy masses energy
feedback from supernovae (SNe) becomes ineffective at suppressing star
formation while cold flows become inefficient at carrying a supply of
fresh gas necessary to regrow a stellar disk \citep{dekel06}. Combined
with the relative higher frequency of major mergers for massive
galaxies \citep{guo08} this framework leads to the build up of a
larger stellar spheroid and disfavors the quick rebuilding of stellar
disks in massive systems.

\begin{figure*}
 \includegraphics[width=14cm]{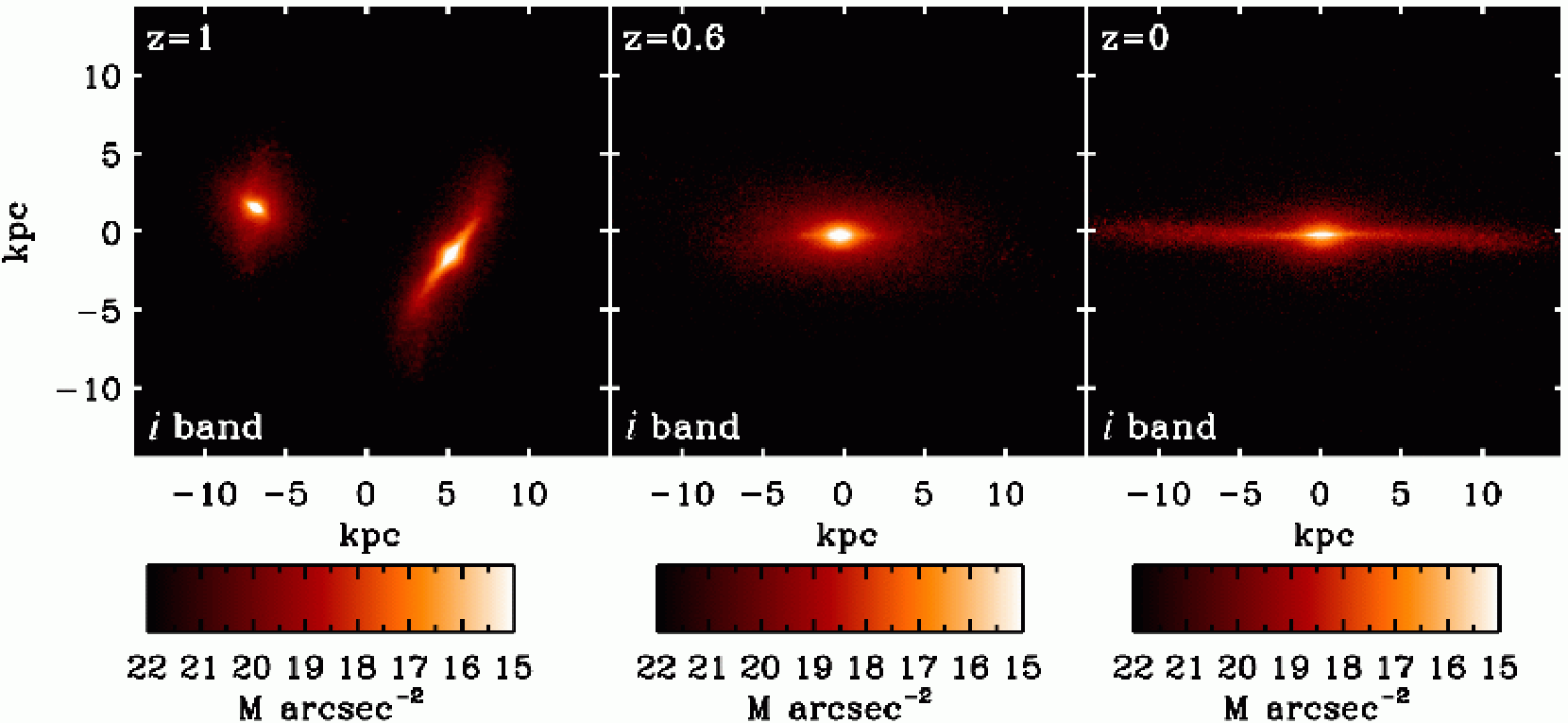}
 \caption{  {\bf SDSS rest frame, unreddened $i$-band surface brightness images of the merging
   system at three different stages}. Left (panel a): z=1.0 close to the
   first pericentric passage. Center (panel b): the merger remnant at z$=$0.6,
   1Gyr after the stellar cores of the two galaxies have coalesced.
   Right (panel c): the system observed at the present day: the halo has faded
   considerably while the galaxy has grown an extended thin disk. The
   $i$-band disk scale length R$_d$ is 7.2 kpc and the light
   ratio of the kinematically identified Bulge and Disk components (B/D) is
   0.49, while a GALFIT 2D decomposition gives a B/D = 0.65.  }
\label{fig1}
\end{figure*}

Only recently have numerical simulations been able to directly compare
the observable quantities of the outputs with real data
\citep{jonsson06,chakrabarti07,lotz08,covington08,rocha08}, rather
than simply predicting the mass distribution of the simulated stellar
systems.  This is a crucial point as the stellar spheroids will fade
drastically after a major merger, while reforming disks will contain a
large fraction of younger and brighter stars. Different mass to light
(M/L) ratios for the two components will skew the observed photometric
light ratios compared to the underlying stellar masses.

Despite all of this progress, the effect of mergers on the existing
disk components, and of feedback and cold flows on the regrowth of
stellar disks have not been studied in detail in fully cosmological
simulations of major mergers. Here we present results from a fully
cosmological, smooth particle hydrodynamic (SPH) simulation where late
gas accretion plays a major role in the rapid regrowth of a dominant
stellar disk in an L$\star$ galaxy after a z $=$0.8 major merger. In
this work we focus on the physical processes that drive the regrowth
of the disk as identified by its kinematical properties, but we also
measure the structural properties of the simulated galaxy based on the
light distribution, i.e. in a way closely comparable to
observations. This result helps solve the apparent contradiction that
strong interactions at relatively low redshifts are common even for DM
halos that are likely to host bright disk galaxies. The paper is
organized as follow: \S 2 discusses the simulations, \S 3 describes
the evolution of the system, \S 4 describes the observational
properties of the merger remnant and \S 5 the assembly of the disk
component. The results are then discussed in \S 6.

\section[]{Description of the Simulation}

The simulation described in this paper is part of a campaign of high
resolution simulations aimed at studying the formation of field
galaxies in a WMAP3 cosmology ($\Omega_0$=0.24, $\Lambda$=0.76,
h=0.73, $\sigma_8$=0.77, {\bf $\Omega_b$=0.042}).  Our sample of
  halos has been selected from low resolution volumes of size 50 and
  25 Mpc (the latter for dwarf size halos) using the ``zoom-in''
  tecnique to ensure a proper treatment of tidal torques
  \citep{katz93}.  The sample covers two magnitudes in total mass
(from 2$\times$10$^{10}$ to 2$\times$10$^{12}$ M$_\odot$) sampling a
representative range in halo spins and epochs of last major
merger. The galaxy described in this paper has a total mass of
7$\times$10$^{11} $M$\odot$, spin $\lambda$ = 0.04 (as defined in
\cite{bullock01}) and a last major merger at z$=$0.8. The environment
density at z$=$0 is typical of field halos with an over density
$\delta \rho$/$\rho$ = 0.1 (0.2) measured over a sphere of radius 3
(8) h$^{-1}$ Mpc. The final galaxy has 650k, 320k and 1.8m DM, gas and
star particles within the virial radius of the galaxy at the present
time, with particle masses: 1.01$\times$10$^6$, 2.13$\times$10$^5$ and
6.4$\times$10$^4$ M$_{\odot}$ for each DM, gas and star particle at
the moment of formation.  The force spline softening is 0.3 kpc. The
minimum smoothing length for gas particles is 0.1 times the force
softening.  The simulations were performed with the N-body SPH code
GASOLINE \citep{wadsley04} with a force accuracy criterion of $\theta$
= 0.725, a time step accuracy of $\eta$ = 0.195 and a Courant condition
of $\eta_C$ = 0.4.  The adopted star formation and SN schemes have been
described in detail in \cite{stinson06} and \cite{g07}.  Our
``blastwave'' feedback scheme is implemented by releasing energy from
SN into gas surrounding young star particles.  The affected gas has
its cooling shut off for a time scale associated with the Sedov
solution of the blastwave equation, which is set by the local density
and temperature of the gas and the amount of energy involved. At the
resolution of this study this translates into regions of $\sim$
0.2-0.4 kpc in radius being heated by SN feedback and having their
cooling shut off for 10-30 million years. The effect is to regulate
star formation in the disks of massive galaxies and to lower the star
formation efficiency in galaxies with circular velocity in the 50 $<$
V$_c$ $<$150 km/sec range \citep{brooks07}.  At even smaller halo
  masses (V$_c$ $<$ 20-40 km/sec, with V$_c$ = sqrt(G $\times$ M/r))
  the collapse of baryons is partially suppressed by the cosmic UV
  field \citep{hoeft06,g07}, here modeled following
  \cite{haardtmadau96}.  The simulation applied a correction to the UV
  flux for self shielding of dense gas as introduced in
  \citet{pontzen08} and low temperature metal cooling
  \citep{wadsley08}. It is important to note that the only two free
parameters in the SN feedback scheme (the star formation efficiency
and the fraction of SN energy coupled to the ISM) have been fixed to
reproduce the properties of present day galaxies  (star formation
  rates, Schmidt law, cold gas turbulence, disk thickness) over a
  range of masses \citep{g07}.  Without further adjustments this
  scheme has been proven to reproduce the relation between metallicity
  and stellar mass \citep{brooks07,maiolino08} and the abundance of
  Damped Lyman $\alpha$ (DLA) systems at z=3 \citep{pontzen08}.
However, even at this resolution the central regions (r$<$1-2 kpc) of
galaxies remain still partially unresolved and likely form bulges that
are too concentrated \citep{g08,mayer08}.  In this respect the mass of
the bulge component of our simulated galaxy has to be considered an
upper limit imposed by current resolution limits, especially at high
z, where the number of resolution elements per galaxy is less.
 
To properly compare the outputs from the simulation to real galaxies
and make accurate estimates of the {\it observable} properties of
galaxies, we used the Monte Carlo radiation transfer code {\it{SUNRISE}} 
\citep{jonsson06} to generate artificial optical images (see Figure 1)
and spectral energy distributions (SEDs) of the outputs of our run.
{\it{SUNRISE}} allows us to measure the dust reprocessed SED
of every resolution element of our simulated galaxies, from the far UV
to the far IR, with a fully 3D treatment of radiative transfer. Filters
mimicking those of the SDSS survey \citep{adelman06} are used to
create mock observations. 

To measure the rotational velocity of the remnant and other galaxies
in the sample in a way comparable with observations we used the
spatial and kinematic distribution of cold gas in the disk of our
simulated galaxy. Using the HI fraction calculated by GASOLINE, we
determined the HI line width, ($W_{20}$), by finding the width of the
HI velocity distribution at 20\% of the peak.  The value of $W_{20}$/2
is then used as a measure of the galaxy rotation velocity at different
redshifts. This measurements reflects the mass weighted
position-velocity distribution of cold gas inside the central region
of the galaxy and in  observed bright galaxies it is usually associated
with the peak rotational velocity.  Because the mass inside the
central few kpc is likely overestimated owing to resolution effects
\citep{mayer08}, $W_{20}$/2 provides a slightly larger measurement of
the rotational velocity than that obtained from the rotational speed
of young stars at 2.2 or 3.5 disk scale lengths (typically 10--20kpc
for bright galaxies), and thus is a useful upper limit for a
comparison with real data.

\begin{table}{}
\centering
\caption{Merger remnant properties at z$=$0.8 (shortly after the
  merger) and at the present time.  $R_d$ is the disk scale length and
  B/D is the bulge to disk light ratio measured for the kinematically
  identified components.  (B/D ratios in parentheses are measured
  using a 2D photometric decomposition with GALFIT of the face on
  projection of the light distribution).  Total magnitudes and colors have been 
  measured in SDSS filters, including the effects of dust (disk
  inclination 45deg) and in the AB system.  }
  \begin{tabular}{@{}llr@{}}
  \hline 
  -  & z=0. & z=0.8 \\
 \hline
 Tot Mag $i$ band  &  -22.3   & -22.7 \\
 R$_d$ i band (kpc)& 7.2 &  4 \\
 B/D $i$ band & 0.49 (0.65) & 1.1 (1.4) \\
  B/D (stellar mass) & 0.87 & 1.16   \\
 $g-r$  & 0.5  & 0.4 \\
 SFR  & 2.2 M$_{\odot}$/yr & 6 M$_{\odot}$/yr\\
 disk stellar mass       & 3.24$\times$10$^{10}$ & 2.07$\times$10$^{10}$ \\

\hline
\end{tabular}
\end{table}

\begin{figure}
 \includegraphics[width=8cm]{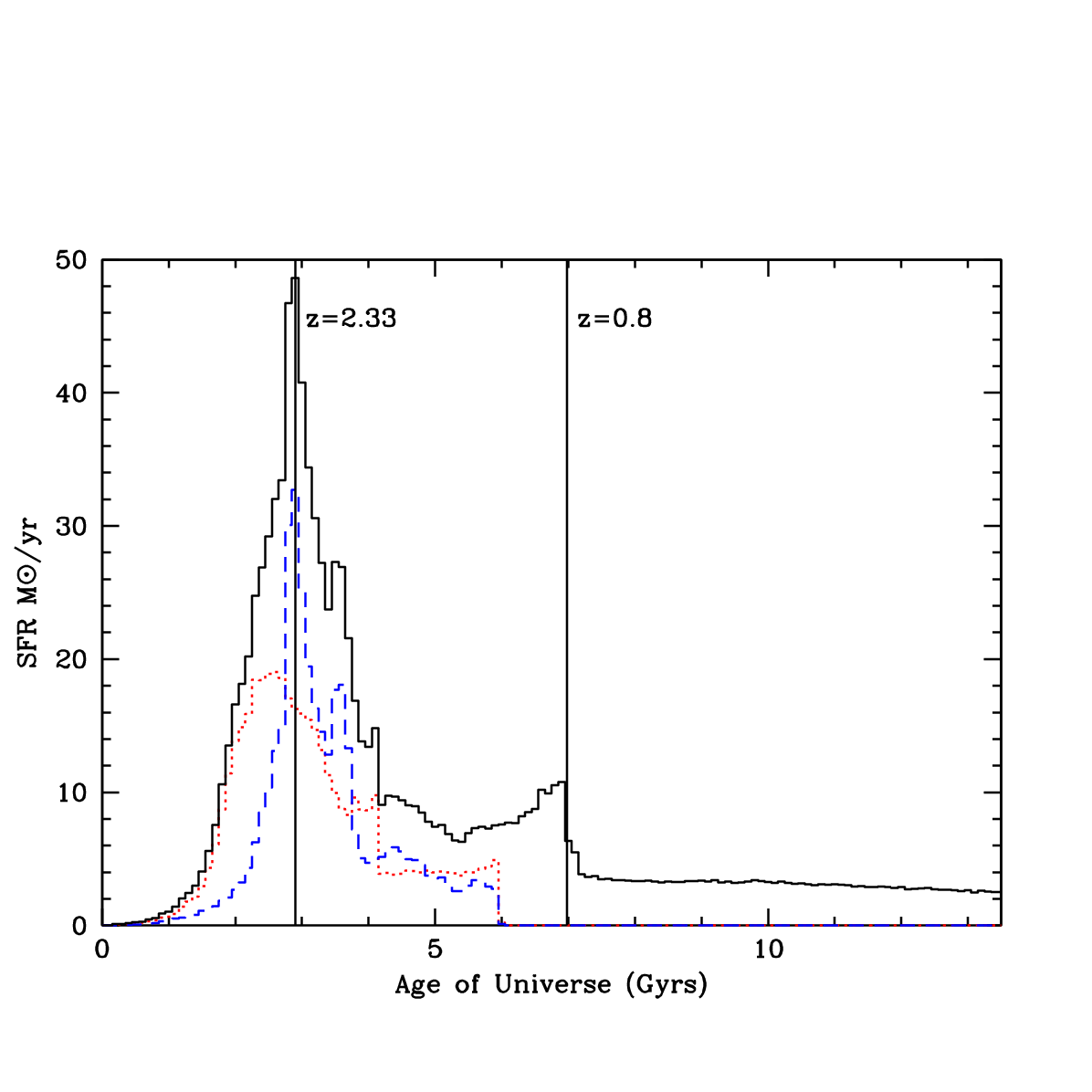}
 \caption{SFR vs time for the last major merger progenitors (red dotted and blue dashed) 
and for the whole galaxy (solid) . The peaks of the dotted and dashed curves correspond 
to the two major  mergers between the original four  progenitors.}
\label{fig2a.ps}
\end{figure}

\begin{figure}
 \includegraphics[width=9cm]{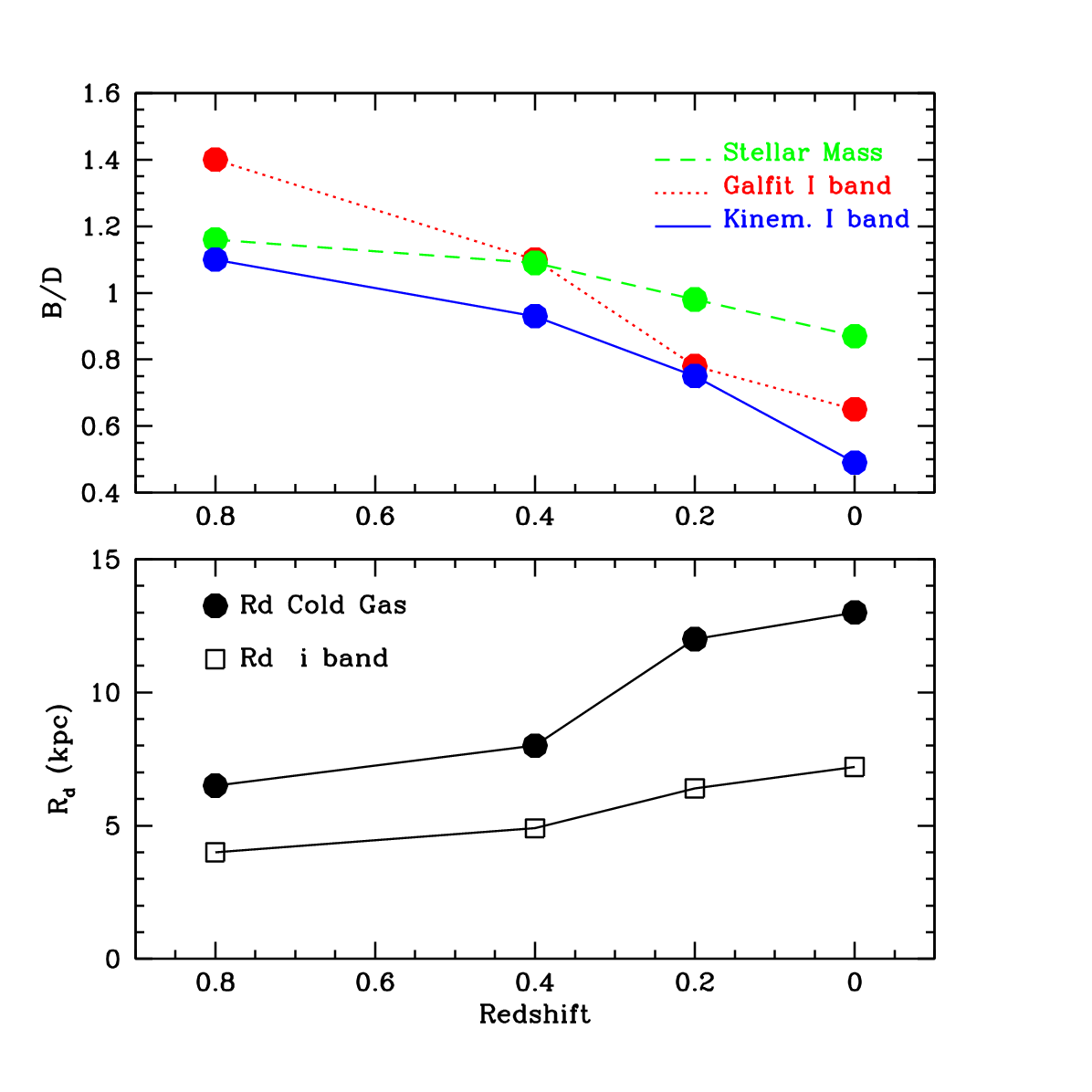}
 \caption{Lower Panel: Time evolution of the cold disk gas (T$<$4$\times$10$^4$ K) scale
   length and the stellar disk scale length (measured in the unreddened $i$ band)
   of the final merger remnant.  Upper Panel: The B/D ratio for the
   final remnant as a function of redshift: dashed: stellar
   masses. Dotted: B/D light ratio (unreddened $i$ band) with a GALFIT
   decomposition based on the 2D, face on light profile. Continuous:
   B/D light ratio (unreddened $i$ band) of the kinematically defined disk and
   bulge components.}
\label{fig3b}
\end{figure}

\section{Dynamical and Photometric Evolution of the Merging System}

The galaxy studied here was singled out for its particularly
interesting assembly history, as the build up of its stellar component
involves numerous major mergers, seemingly a hostile environment to
build a significant stellar disk.  At z$=$3 the four most massive
progenitors form a hierarchy of two binary systems roughly aligned on
the same large scale filamentary structure.  Each halo has a total
mass $\sim$ 7$\times$10$^{10}$ M$_{\odot}$ and has formed a
rotationally supported stellar disk fed by strong cold flows, typical
of galaxies at that redshift \citep{brooks08}.  In this work we define
as ``cold flow'' gas that has never been shocked to 3/8 the virial
temperature of the parent halo \citep{keres05, brooks08}.  The average
cold (T$<$4$\times$10$^4$ K) gas fraction in  the  disks  of the four 
  galaxies is 25\% (defined as the fraction of total baryons in the
disk). Both pairs merge by z$=$2. Just before the mergers, the
four progenitors have rest frame $B$ magnitudes in the range -21.1 --
-20.2 (-21.5 -- -20.6 in the r band) and rest frame $g-r$ colors
around 0.3 -- 0.4 (unless specified all global magnitudes and colors
in this work are in the rest frame AB system and include the effects
of dust reddening measured at a 45 deg inclination). These two early
mergers are then ``wet'' i.e. between galaxies with blue colors as
defined in \cite{lin08}.

Both merger remnants quickly reform extended  gas disks from a
combination of freshly accreted gas and gas already in the progenitors
disks that was not turned into stars during the merger.  For each
merger the star formation history (SFH) peaked at 18 and 32
M$_{\odot}$/yr respectively (Figure 2). At z$=$1.6 (2 Gyrs before the
final merger) the two disk galaxies formed from the early mergers have
again very similar magnitudes: -21.6 (-22.1) and -21.3 (-21.8) in the
B (r) band respectively, close to the B band L$\star$ at that z
\citep{marchesini07a}. Their $g-r$ color $=$ 0.4, makes them bluer
than most galaxies of similar brightness at the present time
\citep{lin08}.

The final major merger of these two progenitor galaxies begins at
around z$=$1 when their dark matter halos, flowing along the same
filamentary structure, first overlap.  The galaxies plunge in on
fairly radial orbits, with the internal spins of the two disks roughly
aligned with the orbital angular momentum vector (Figure 1a). Note
that this is not necessarily a configuration favorable to the survival
of gaseous disks, as noted by \cite{hopkins08}. After two close
passages, the two galaxies coalesce by $\sim$ z=0.8, i.e., 1 Gyr after
the merger commenced. The mass ratio of the merging halos is 1.2:1
while the $i$ band brightness ratio is 1.6:1.  During the merger, the
global star formation rate of the system is enhanced and peaks at 11
M$_{\odot}$ with subsequent star formation rates in the remnant
dropping rapidly by almost factor of three to $\sim 4$ M$_{\odot}$/yr
(Figure 2).  The moderate star formation (SF) enhancement is
consistent with estimates for interacting systems in the same redshift
range \citep{jogee08}.  Given the galaxies' properties outlined above,
even this final merger is then clearly identified as ``wet''. However,
the disks of the two galaxies have gas fractions around 20\%, so they
are only relatively gas-rich compared to the present day population of
galaxies of similar brightness \citep{garcia08} and have equal or
lower gas abundance as z $\sim$ 2 galaxies \citep{erb06}.

Once again, following the final major merger a  gas disk rapidly
regrows and star formation is mainly concentrated in the reforming
 stellar disk. The galaxy remains relatively unperturbed after
z$=$0.8. Shortly after the merger the spheroidal component of the
newly formed galaxy dominates the light distribution (Figure 1b),
although a disky component is already visible. The $g-r$ color is 0.5
and remains stable to the present time.  We verified that the disky
component at z=0.8 is indeed associated with a thick stellar disk
supported by rotation.  By z$=$0 the galaxy has regrown an extended
thin disk,  and while about 50\% of the baryons in the halo have
  been turned into stars, the spheroid has faded considerably.  The
halo has faded in the reddened B band by 0.8 mag to B = -21, due to
the aging of the spheroidal component. The disk clearly dominates the
light distribution in the SDSS $i$ and bluer bands (Figure 1c).

These results  show that at all stages the progenitors of the final
galaxy can be identified with a normal population of moderately gas
rich disk galaxies with colors and SF rates comparable with the galaxy
population observed in the 3$>$z$>$1 range. It is also important to
emphasize that the cold gas fraction in all the disk progenitors
suggests that idealized initial conditions with disk gas fraction as
high as 50\% are not a necessary condition to have ``wet'' mergers and
rapid regrowth of stellar disks.  As discussed in $\S$5 the
fundamental ingredient for disk regrowth and blue colors is the fast
refueling of gas from the hot halos and the surrounding cosmic web,
highlighting the necessity of a fully cosmological approach to the
problem of the abundance of galaxy disks$^1$.\footnotetext[1]{Movie at www.astro.washington.edu/fabio/movies/Merger.mpg}

\begin{figure}
 \includegraphics[width=8cm]{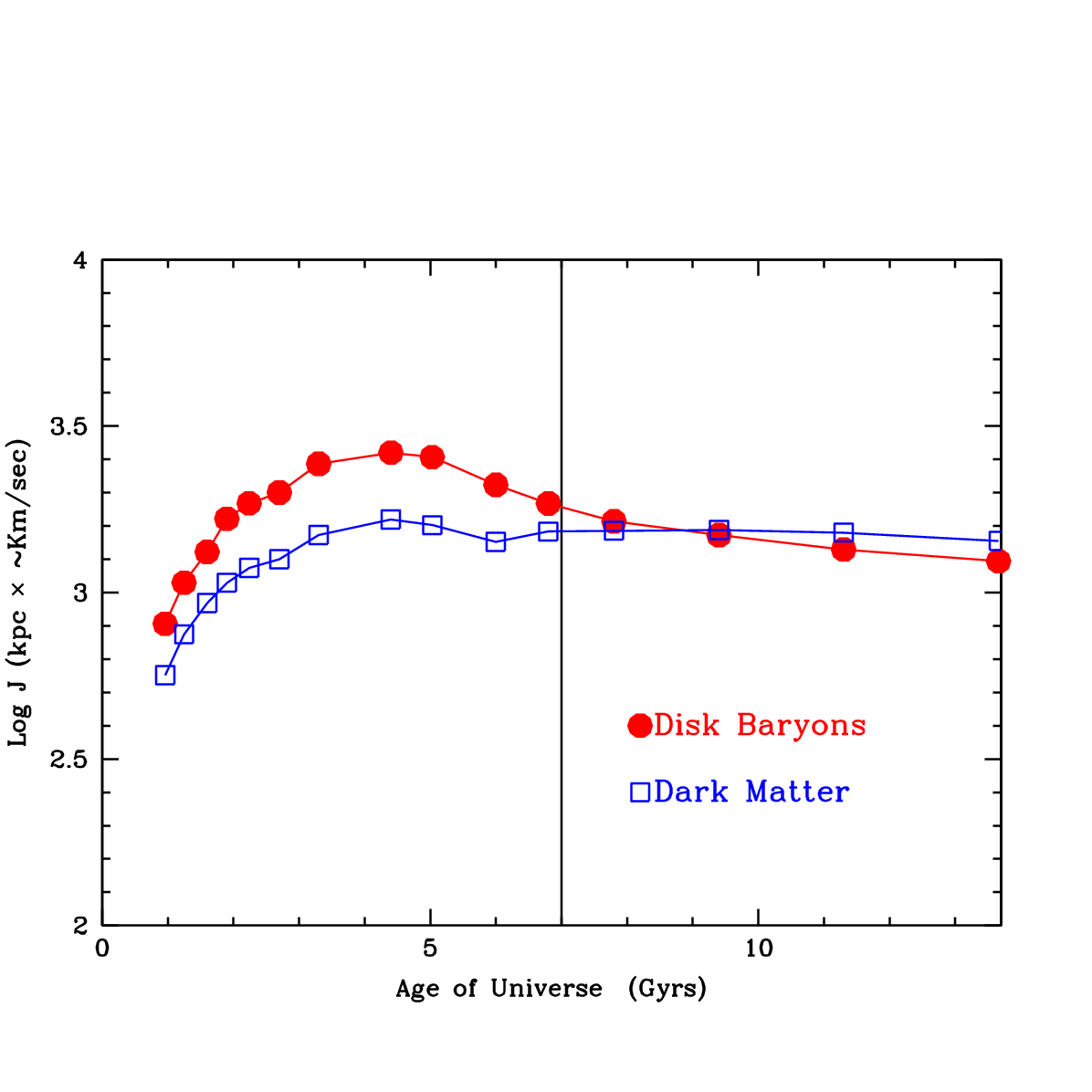}
 \caption{The time evolution of the angular momentum content per unit
   mass of the kinematically identified z=0 baryonic disk (gas and
   stars) and the dark matter of the merger remnant. The reference
   frame is defined by the center of mass of the selected particles.
   Disk material conserved most of the angular
   momentum gained by z=1.5. The vertical line marks the
   epoch when the halos and the baryonic cores merge during the last
   major merger. }
\label{fig4}
\end{figure}

\section{Properties of the Merger Remnant from z$=$0.8 to the Present}

In this section we decompose our simulated galaxy into its kinematic
components and ``observe'' some of its properties at the present day
(they are summarized and compared with those of the galaxy at z$=$0.8
in Table 1). Does the galaxy have the photometric and kinematic
properties of disk dominated galaxies?

At the present time the disk clearly dominates the light distribution
even in the relatively red $i$ band (Figure 1c) The galaxy has a total
magnitude is $i$ $=$-22.3 (-22.4 unreddened) mag and global reddened
color $g-r =$0.5, consistent with those of luminous present day
disk galaxies \citep{lin08}. 

To measure its morphology in a quantitative way the different
components of the galaxy were first identified using their kinematic
and spatial information and classified as bulge, halo and disk. This
is a crucial step to relate each galaxy component to its physical
origin.  First the disk plane is defined using the cold gas in the
  central few kpc of the galaxy, then disk stars are defined as stars
whose specific angular momentum perpendicular to the disk plane
(j$_z$) is a significant fraction of the maximum angular momentum of a
circular orbit with the same binding energy (j$_c$), i.e j$_z$/j$_c$
$>$ 0.8. Particles on circular but  inclined orbits (more than 30 deg) are
excluded.  Bulge and halo stars were then identified based on their
radial orbits and their binding energy (bulge stars being more bound).
The energy separation criteria between halo and bulge stars
corresponds to the radius at which the spheroid mass profile changes
slope (halo stars having a shallower profile than bulge stars) and in
our simulated galaxy sample separates an older and metal poor
population (the halo) from bulge stars that are more metal rich. Halo
stars contribute a fraction ($\sim$15\%) of the total stellar mass
within the virial radius of the galaxy, but the halo central density
is two orders of magnitude lower than that of the bulge. Hence the
details of the bulge/halo decomposition do not change our
conclusions. At z$=$0 the kinematically identified disk, bulge and
halo stellar masses are respectively: 3.4, 2.7 and
1.$\times$10$^{10}$M$_{\odot}$.  We then imaged each separate
component using {\it{SUNRISE}} and measured their structural
parameters using the unreddened images. We focused on a structural
analysis of the unreddened components, avoiding the additional layer
of complexity given by the details of the dust distribution, which
will be explored in future papers with a larger number of galaxies.
However, we have verified that our findings do not change if the
reddened images are used instead.

How and when did the disk reform after the last major merger at
z$=$0.8?  The stellar disk and bulge components were identified at
different redshifts after the last major merger event.  To better
understand the B/D ratio evolution of the merger remnant we measured
B/D in three different ways.  We used the kinematic decomposition to
find a) the stellar mass ratio of the bulge and disk components and b)
their relative flux ratio in the $i$ unreddened band. Then we analyzed
the unreddened, face-on 2D light distribution created by {\it SUNRISE}
using all the galaxy star particles (including halo particles) with
GALFIT \citep{peng02} to find c) the B/D ratio as determined by a fit
to the surface brightness profile.  Figure 3 shows how the $i$ band
disk scale length and the B/D ratio evolve with time. Shortly after
the merger event the disk component is already visible edge-on, then
the bulge component fades relative to the disk and the disk becomes
more extended. At z$<$0.4, or about 3.5 Gyrs after the merger, the
disk dominates both in terms of the light contribution and the B/D
ratio decreases further by z$=$0.  All measurements agree on the same
trend of B/D decreasing with time.  At the present time the disk
extends almost far as 20 kpc in radius from the galaxy center and the
stellar disk scale length R$_d$ is 7.8 and 7.2 kpc in the B and $i$
bands, respectively (Figure 3), consistent with observations of real
galaxies that show larger R$_d$ in bluer bands.  Smaller B/D ratios
are obtained using the light distribution, more sensitive to the
younger ages of disk stars. At z$=$0 the B/D stellar mass ratio of the
kinematically defined components is 0.87, but the unreddened $i$ band
light ratio is only 0.49.  GALFIT shows the steepest trend with age,
and shortly after the merger it underestimates the disk component, if
by less than 20\%.  GALFIT gives a fairly precise estimate of the
light weighted B/D ratio when the stellar disk becomes dominant.

Soon after the merger the cold gas disk increases its size by nearly a
factor of two as the new infalling gas settles on high angular
momentum orbits. The angular momentum of the present day disk baryons
is mostly acquired at high z (Figure 4) as predicted in analytical
models \citep{quinn92} with a fraction of it transferred to the DM
halo during the last major merger.  As this gas is gradually converted
into stars the stellar disk also grows in size, however the cold gas
disk remains significantly more extended than the stellar one.
Shortly after the merger the (unreddened $i$ band) disk scale length
is 4 kpc, at the present time it is almost twice as large, with
only minor warping (Figure 7).  Most likely due to the late assembly
of its younger component, this is a fairly extended disk for galaxies
of this mass, more extended than many disks formed in cosmological
simulations where the assembly history of the galaxy was more
``quiet''. Encouragingly, we verified that the structural properties
of the bulge and disk do not change much if they are measured using
GALFIT on the global unreddened 2D light distribution, i.e. without a
prior knowledge of the kinematic decomposition.  With GALFIT the $i$
band R$_d$ is 6.8 kpc, a 5\% difference.  GALFIT finds systematically
larger B/D ratios, but only by 20\% or less. However, B/D ratios
decrease if the GALFIT fitting is done on dust reddened images.  These
B/D ratio and R$_d$ are quite typical of bright Sa and Sb galaxies
that typically have dust corrected B/D $\sim 0.5$ and red R$_d$ $\sim$
3--7 kpc \citep{graham08,driver07c,benson07}.

To show how the simulated galaxy of this study relates to the general
population of real disk galaxies on the Tully Fisher (TF) relation, we
compared its HI velocity width $W_{20}$/2 (238 km/sec) and total
rest-frame unreddened i-band relation with  the disk dominated
galaxies in our simulated sample and with the sample of disk galaxies
described in \cite{geha06} and references therein (Figure 5).  We find
that the agreement between the observed TF and our set of simulations
is quite good, as the combination of high resolution and the feedback
adopted in our simulations yields a good match to real galaxies over a
wide range of magnitudes and circular velocities.  The merger remnant
object of this study lies well within the observed scatter of both the
observed and simulated samples, confirming that its structural
properties are similar to those of typical bright spiral galaxies.  We
will present the scaling properties of the full data-set of simulated
galaxies in a forthcoming paper.  An interesting feature of the Tully
Fisher plot is the evolution of the remnant on the TF plane: the very
limited growth of the bulge stellar mass (only a few \%) and the
inside out growth of the disk ensures that the amount of mass within
the central region of the galaxy does not change, hence $W_{20}$/2
does not evolve strongly, while the overall fading of the stellar
components makes the galaxy dimmer by less than half a magnitude
between redshift 0.8 and 0 (see Table 1). Even if the assembly history
of our galaxy is not typical of galaxies of similar total mass, this
result is consistent with observations that find a small evolution in
the observed galaxy TF relation up to z$=$1 \citep{conselice05} along
with strong size evolution \citep{trujillo07}.  This result is not
trivial and we plan to extend this analysis to our full sample of
simulations in a future work.

 We also verified that the metallicity of the cold gas is 
consistent with the observed stellar mass - metallicity relation 
\citep[8.5 $<$ 12+log(O/H) $<$ 8.9, depending on the aperture used,][]{tremonti04,brooks07}.
An average metallicity consistent with real galaxies is an important test
of the realism of this simulation and makes the estimates of the
galaxy colors of the final remnant and its progenitors more robust.
Finally, the satellite system of the remnant includes 11 resolved
luminous satellites within the virial radius of 230 kpc. The faintest
has AB B mag = -8.7, the brightest -17.9. By tracking the satellites
through different outputs after the merger event we verified that the
galaxy disk undergoes several fly-bys by small dark satellites, but no
significant accretion of luminous satellites after the final
merger. Many faint satellites have undergone severe tidal stripping of
both their DM halos and of their stellar component.
\begin{figure}
 \includegraphics[width=9cm]{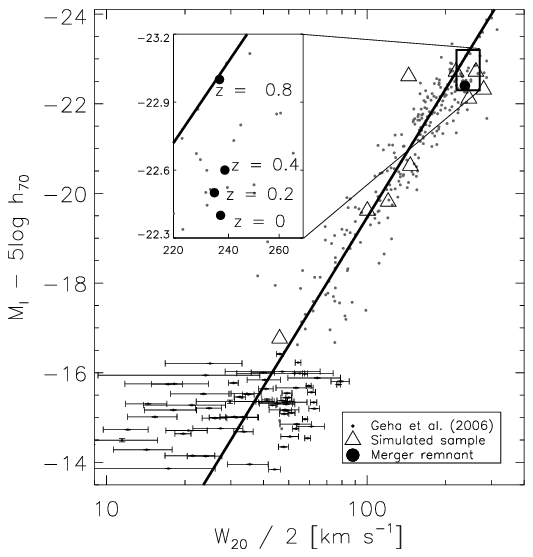}
 \caption{The location of the simulated galaxy in this work on the
   present  day ``Tully Fisher'' relation, i.e., the unreddened $i$ 
   band vs $W_{20}$/2.
   Grey dots: Galaxies from a compilation of observational data
 \citep{geha06}. Triangles: other disk galaxies from our
   simulated sample at similar resolution. Filled dot in main panel:
   the galaxy of this study at z=0.  Filled dots
   in upper left insert: different snapshots of the merger remnant
   from z=0.8 to the present. Circular velocities are the $W_{20}$/2
   velocity widths from the HI distribution of the simulated
   galaxies. The insert shows the weak redshift evolution of the
   merger remnant on the TF plot.}
\label{fig5}
\end{figure}
This analysis quantifies the dramatic regrowth of the disk component
and how, coupled with the fading of the bulge and halo components it
leads to the formation of a galaxy dominated by an extended disk.
It also highlights the difference between evaluating a galaxy
morphological type using the mass distribution compared to the light
distribution. It is encouraging however, that similar trends (growing
disk size and decreasing B/D ratios) are recovered using complementary
techniques, as GALFIT provides a decomposition into bulge and disk
components quite similar to that obtained using the full spatial and
kinematic information. These results support the notion that the
observed B/D ratios in  bright galaxies are indeed representative
of the underlying dynamical disk and spheroidal components.

\section{The Assembly of the galaxy disk}

\begin{figure}
 \includegraphics[width=8cm]{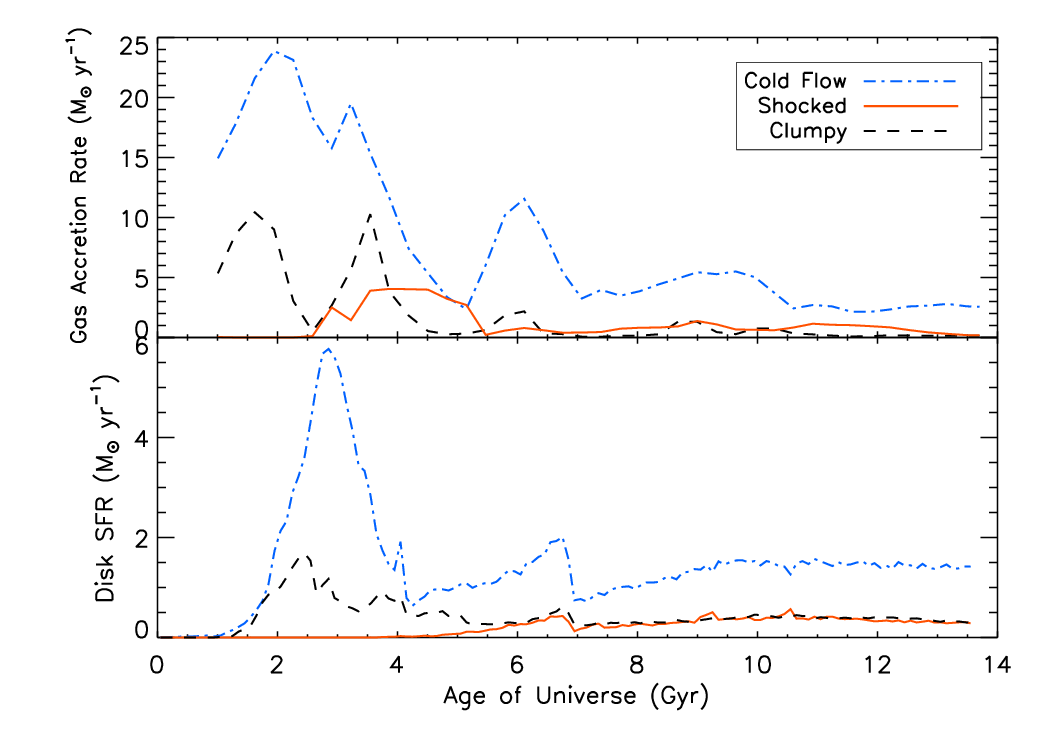}
 \caption{ Top Panel: The total gas accretion history of the merger
   remnant, divided by the thermodynamical history of the gas. (at
   high z the lines are the sum of the contributions from individual
   progenitors). Bottom Panel: The SFH of disk star particles separated
   by the history of their parent gas particles.  The disk stars are
   kinematically identified at z$=$0.  Dashed: from clumpy gas
   accretion, dot dashed: from unshocked gas (or ``cold flows''),
   solid: shocked gas. A significant fraction of stars formed at high
   z and were identified as part of the disk even after two major
   mergers. At the present time they form the thick disk component
   (see also Figure 7).   Star formation in the disk is partially
   disrupted during the last major merger, and stars formed from gas
   cooled from the hot halo form a significant fraction of the younger
   disk component.}
\label{fig2b.ps}
\end{figure}

 A detailed analysis of the gas accretion history was
performed by tracking backward every gas particle that was ever within 
the virial radius of the simulated galaxy and its progenitors.
Every star particle comes from a gas progenitor which is uniquely 
identified at the moment of the star particle formation.  The 
thermodynamical history of each gas particle progenitor was then 
studied.  This temperature history was done for each of the original 
four progenitor galaxies, then again for the resulting two binary 
galaxies, and finally for the merger remnant.  
Gas particles were then classified as a) ``cold
flows'' if it never belonged to a progenitor halo before being
accreted onto  one of the progenitor galaxies or the merger 
remnant and it never shocked to 3/8 the virial temperature 
of the galaxy, b) ``shocked'' if the gas was first shocked to 
$>$ 3/8 the virial temperature of the main halo and then cooled onto 
the disk, and c) ``clumpy'' if the gas particle was assigned to
a subhalo before becoming part of the disk (i.e., a halo other than 
one of the four main progenitors and their subsequent merger remnants).

Figure 6 shows (top panel) the total gas accretion rates to progenitor
galaxies and merger remnant, and (bottom panel) the SFH of the
resulting disk (as identified at z$=$0 based on the stellar
kinematics), separated by the different histories of each gas particle
progenitor.  It is remarkable that a large fraction of the disk formed
at redshift one or earlier, when the galaxy had not undergone its last
major merger yet. This result supports the suggestion that major
mergers do not always completely destroy the pre-existing disks
\citep{hopkins08}, although they heat them quite substantially, making
them thicker. Cold flows build the largest fraction of the disk up to
low redshift. This analysis shows how the build up of the disk of
bright galaxies differs from the simple picture of gas cooling from
the hot halo and assembling into a disk; only at later times does gas
cooling from the hot halo phase plays a role. From Figure 6 it is
clear that cold flows dominate the early gas accretion history of the
progenitor galaxies, and is the dominant source of SF in the galaxy
disk.  Because SN feedback regulates SF in the disk, a cold gas
reservoir is able to develop that sustains SF until the present time.
The SF rate drops significantly when the two progenitor disk galaxies
merge at z$\sim$1, After this time, gas cooling from the hot halo
becomes more important, resulting in 30\% of the disk stars formed in
the last 2 Gyrs formed from gas labeled as shocked. However, only 10\%
of the total amount of stars in the disk at the present time have
formed from gas that was previously heated to the virial temperature
of the main progenitor and almost 40\% of the disk stars after z$=$0.8
formed from cold flows accreted after the last major merger.  We find
that this result is quite typical for disk galaxies of total mass $<$
10$^{12}$ M$\odot$ \citep{brooks08} stressing the role of cold flows
in building early disks and of ``shocked'' gas in rebuilding some of
the young, and hence bluer, more metal rich and brighter part of the
disk. These effects will have to be carefully modeled in semi
analytical models of galaxy formation.
\begin{figure}
 \includegraphics[width=8cm]{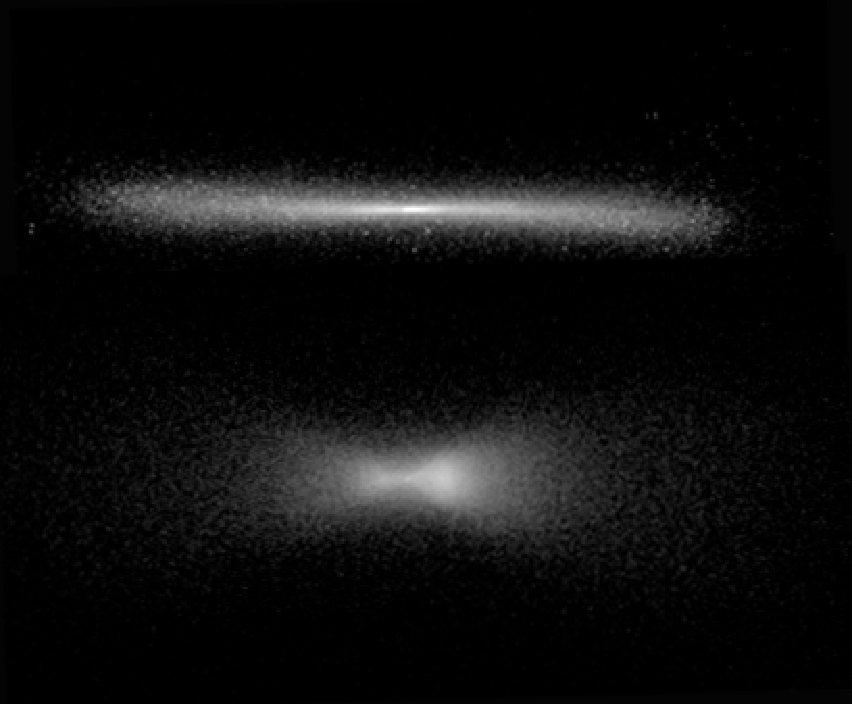}
 \caption{The $i$ band (unreddened) edge on image of the galaxy disk
   at z$=$0. Upper Panel: light distribution from stars formed after
   the last major merger. Lower panel: the light distribution from
   stars formed before the last major merger, but still identified as
   disk based on their angular momentum. The image is 40 kpc across
   The relative brightness of the two components is not to scale.  }
\label{fig7}
\end{figure}
Figure 1 shows that as the disk grows in mass, most of
the stars are added to a thin and dynamically cold part of it. To
quantitatively evaluate this process we separated the stars previously
dynamically identified as disk into two populations: those that formed
before the z$=$ 0.8 merger and those formed after, i.e in the last
$\sim$ 6 Gyrs.  Figure 7 shows the edge-on $i$ band surface brightness
image (unreddened) of the two populations. Their properties are
strikingly different: the younger stars form a thin exponential disk
with $i$ band scale length of 8.4 kpc (measured using stars in the
2-13 kpc range) and scale height of just 0.5 kpc (measured as a
Gaussian distribution at a radius of 8.5 kpc). This scale height is
likely as small as allowed by the force softening of 0.3 kpc.  The
older component, while being comparable in mass, is 1.3 mag fainter in
the $i$ band and 2.2 mag in the B band. It also has a much shorter
disk scale length, 3.1 kpc, and is considerably thicker, 2.7 kpc,
again at an 8.5 kpc radius. The flaring in the old disk is due to the
kinematic selection criteria, that exclude particles on circular
orbits that are highly inclined from the disk plane.

While beyond the scope of this paper, we speculate that the mix of the
two components could easily be identified as a thick and thin disk
components if observed using typical observational techniques
\citep{yoachim06}. Results from this simulation support the notion
that thick disks are formed during major mergers and the early
assembly of galaxy disks \citep{brook04}. Furthermore, the last major
merger should leave a clear signature in the age distribution of thick
 disk (which will be older than the merger) and thin disk stars 
(mostly formed after  the merger).

In this realization no significant component of the stellar disk
formed by accretion of stars in smaller satellites through minor
mergers.  While a substantial galaxy to galaxy scatter is expected,
this is consistent with results from our larger sample of simulated
disk galaxies \citep{brooks08} and, compared with the existing
literature, a consequence of the smaller stellar masses of the galaxy
satellites resulting from the realistic feedback implementation
\citep{g07}.  We verified that the bulge grows only modestly in mass
from z$=$0.8 to z$=$0 and that no strong bar instabilities form after
the last major merger. Similarly the halo component shows minimal mass
growth during the same time period.  A close examination of the time
evolution of the system shows several passages and disruption of small
dark satellites, but they do not prevent the system from reforming a
thin stellar disk.

\section{Conclusions}

We have analyzed an SPH, fully cosmological simulation of the
evolution of a galaxy that grows an extended thin stellar disk after a
major (1.6:1) merger at z$=$0.8. The disk dominates the light
distribution 3.5 Gyrs after the merger. By the present time the galaxy
shares several properties with the observed population of disk
dominated L$\star$ galaxies.

This result is particularly relevant as the assembly history of the
galaxy's parent DM halo is different from many previously published
studies of cosmological simulations of disk galaxies, as it undergoes
a late major merger, whereas previous works tended to select galaxies
with a relatively quiet merger history. The two progenitors underwent
major mergers themselves, at z $\sim 2$.  The end result of this
simulation strongly contradicts the notion that disk formation
requires a ``quiet'' halo merging history.  In fact, the merging
history of the halo picked for this study has often been considered
hostile to the formation of extended disks at the present time. Our
study provides strong support to the notion that galaxy disks can
reform in a few Gyrs after a gas rich major merger, while a fraction
of the pre-existing stellar disk can survive, even if faded by age and
thickened by the strong interaction.

The results of our analysis are made particularly robust by measuring
the properties of the remnant light distribution rather than just that
of the stellar mass.  This approach is crucial, as it highlights the
effects of fading of the light from older stellar populations and
plays a major role in quantifying the predominance of the newly formed
(and hence bluer and brighter) stellar disk. The main structural
parameters of the galaxy at z$=$0 suggest that it has an Sa or Sb
morphology: reddened $i$ band B/D $\sim$ 0.65, Rd = 7.2 kpc, $g-r$ =
0.5, a $W_{20}$/2 HI velocity width of 238 km/sec.  Moreover, being
able to measure the reddened colors and brightness evolution of the
progenitors allows us a comparison with high-z galaxies,
showing that they are representative of the population of blue, gas
rich and moderately star forming galaxies observed at z$>$1. Verifying
that the progenitors have some of the observed properties of high
redshift galaxies is important, as it makes the present day properties
of the merger remnant more relevant, having been built from realistic
progenitors.

As our simulation includes a full treatment of the cosmological
environment it includes a realistic treatment of a number of
hydrodynamical processes that are necessary for the regrowth of
stellar disks and that have to be simultaneously included: 
  continuous inflow of gas from the cosmic web, stellar feedback, and
  cooling from hot halo gas. This explains the difference between our
results and those of the collisionless simulations of
\cite{purcell08}, which showed significant disk heating due to
interactions and accretion events. Lack of gas infall prevented the
disks in their simulations to regenerate a new thin component (gas
resupply was in fact advocated as a possible solution). In our
simulation the effect of interactions with infalling satellites, both
luminous and dark, is naturally included. However, interactions and
minor mergers are not strong enough to significantly disrupt or
thicken the disk as it reforms from new gaseous material.

Idealized hydrodynamic simulations of binary galaxy mergers
\citep{robertson06,hopkins08} did not include the continued infall of
gas from cold flows and the hot halo. These works pointed out that
without any subsequent accretion/growth onto the disk after the
merger, a cold gas fraction in excess of 50\% would be required in the
disks of the progenitors in order to re-build a disk dominated galaxy
after the merger.  Here we have shown that such high disk gas
fractions are not necessary, as cold flows and cooling from a hot halo
make disk regrowth possible even at low redshifts, when the gas
fraction in the progenitors' disks and the remnant is only $\sim$20\%
at any given time. The analysis of the build up of the disk in our
simulated galaxy highlights the dominant role of cold flows at high
redshift. Cold flows funnel gas to the central disk on a time scale a
few Gyrs shorter than gas that is first shocked to the virial
temperature of the host halo and then cools onto the disk, leading not
only to early disk star formation, but the creation of a large
reservoir of cold gas.  Thus, if a substantial fraction of this cold
gas reservoir survives a merging event, it can provide a faster
accretion rate onto the reforming disk than that available from
cooling the hot gas in galaxy halos alone. Rapid disk reformation will
also be aided if cold gas accretion onto the galaxy halo is still
occurring, while gas cooling from the hot phase forms just 30\% of the
post merger stellar disk. However, this young component is more
evident when the system is observed in bluer bands. It is important
that these processes of disk formation and destruction be carefully
implemented in analytical models that study the properties of large
sample of galaxies.

More work is also needed to make detailed quantitative predictions about
the morphology of galaxies formed in cosmological simulations and to make
the results of our study more general.  The maximum halo mass and
lowest redshift at which mergers can regrow disks are obviously a
function of the feedback efficiency \citep{brook04b,scannapieco08} and
could therefore provide useful qualitative tests of models of SF and
feedback.  In dense environments, such as groups or clusters, the gas
reservoir associated with cold flows and cooling halo gas will likely
be disrupted by tidal forces and ram pressure stripping. Hence the
disk re-growth process should be much less efficient, as expected by
the observed correlation between galaxy morphology and environment
density.  Also, higher resolution cosmological simulations will have
to address the role of secular processes on the detailed structure of
the bulge and thin disk components and their role in setting the B/D
ratio of galaxies \citep{debattista04,genzel08aph,
  weinzirl08aph}. Still, results of the work presented here greatly
alleviate the problem posed for the existence of disk galaxies by the
high likelihood of interactions and mergers for galaxy sized halos at
relatively low z.

\section*{Acknowledgments}

Simulations were run at ARSC, NASA AMES and Texas Supercomputing
Center. FG acknowledges support from a Theodore Dunham grant, HST
GO-1125, NSF ITR grant PHY-0205413 (also supporting TQ), NSF grant
AST-0607819 and NASA ATP NNX08AG84G. CBB acknowledges the support of
the UK's Science \& Technology Facilities Council (ST/F002432/1). PJ
was supported by programs HST-AR-10678 and 10958 and by Spitzer Theory
Grant 30183 from the Jet Propulsion Laboratory.  We acknowledge
discussions with several smart people, among them Avishai Dekel, Mark
Fardal, James Bullock and Phil Hopkins.  FG and AB acknowledge the
hospitality of the Max Planck Institute during the writing of this
paper.

\bibliographystyle{mn2e}
\bibliography{fabio_masterbiblio}

\bsp

\label{lastpage}

\end{document}